\documentclass[pra,aps,twocolumn,amsmath,footinbib]{revtex4-2}
\usepackage{amssymb}
\usepackage{graphicx}
\usepackage{hyperref}

\usepackage{color} 

\definecolor{darkgreen}{rgb}{0,0.7,0}

\usepackage[table]{xcolor}
\definecolor{lightgray}{gray}{0.9}

\usepackage{booktabs}
\newcommand{\ra}[1]{\renewcommand{\arraystretch}{#1}}

\usepackage[caption=false]{subfig}
\newcommand{\phantomsubfloat}[1]{
    {
        \captionsetup[subfigure]{labelformat=empty}
        \subfloat[][]{#1}
    }%
}

\begin{document}

\title{Thermometry of one-dimensional Bose gases with neural networks}

\author{Frederik M{\o}ller\textsuperscript{1}, 
Thomas Schweigler\textsuperscript{1,4},
Mohammadamin Tajik\textsuperscript{1},
Jo{\~a}o Sabino\textsuperscript{1,2,3},
Federica Cataldini\textsuperscript{1},
Si-Cong Ji\textsuperscript{1},
and J\"{o}rg Schmiedmayer\textsuperscript{1}
}

\affiliation{
\textsuperscript{ 1} Vienna Center for Quantum Science and Technology (VCQ), Atominstitut, TU Wien, Vienna, Austria\\
\textsuperscript{ 2} Instituto Superior T\'{e}cnico, Universidade de Lisboa, Portugal\\
\textsuperscript{ 3} Instituto de Telecomunica\c{c}\~oes, Physics of Information and Quantum Technologies Group, Lisbon, Portugal\\
\textsuperscript{ 4} JILA, University of Colorado, Boulder, Colorado 80309-0440, USA
}

\date{\today}

\begin{abstract} 
We design a neural network to extract and process features from absorption images taken of one-dimensional Bose gases in the quasi-condensate regime. Specifically, the network is trained to predict both the temperature of single realizations of the system and the uncertainty thereof. For multiple realizations, the individual predictions can be combined in an estimate of the mean temperature, improving precision. 
We benchmark our model on both simulated and experimentally measured data and compare it to the established method of density ripples thermometry. We find the predictions of the two methods compatible, although the neural network reaches similar precision needing much fewer realizations, thus highlighting the efficiency gain achievable when incorporating neural networks into analysis of data from cold gas experiments.
Further, we study feature maps to reveal which local features of the condensate are extracted by the network and how said features correlate with properties of the system. A similar analysis could be employed to uncover physical relations in more complex systems.
\end{abstract} 

\maketitle 

\section{Introduction}

Machine learning based on neural networks has in many ways revolutionized our ways of processing data~\cite{lecun2015deep}. Recently, the techniques have also started gaining traction in the quantum physics research, where they have been successfully employed for a number of applications~\cite{carrasquilla2017machine, van2017learning, Carleo602, gao2017efficient, rem2019identifying, PhysRevLett.123.230504, bohrdt2019classifying, zhang2019machine, doi:10.1063/1.5048290}. 
Although still in its infancy, the increasing availability of computational power and user friendly machine learning libraries makes it easier than ever before for research groups to explore how they can benefit from these new techniques.
Therefore, establishing and exploring the potential of neural networks in quantum research is highly important.  

Particularly within the field of ultracold quantum gases~\cite{RevModPhys.80.885} could machine learning have a tremendous impact. These quantum many-body systems exhibit complex dynamics and contain a vast amount of information, making theoretical descriptions very challenging. Hence, descriptions often rely on effective or emergent models~\cite{PhysRevA.67.053615, PhysRevX.6.041065, PhysRevLett.117.207201, RevModPhys.77.259, francesco2012conformal, Schweigler2017, PhysRevB.75.174511, PhysRevLett.126.090602}, aiming to reduce the complexity. Neural networks, on the other hand, have demonstrated the ability themselves to find the most efficient representations of complex states~\cite{gao2017efficient, Carleo602, schmitt2021observations}. 

Also in an experimental setting can neural networks be employed to great effect~\cite{Barker_2020, Wigley2016}, particularly in enhancing the readout and analysis of data~\cite{rem2019identifying,Picard_2019}.
The measurement of quantum gases is primarily done using the techniques of absorption or fluorescence imaging, which produce images of the atomic density integrated along the imaging direction. From these images a few physical quantities can be extracted directly, whereas others require fitting with effective theories. One of neural networks greatest strengths is image processing and pattern recognition, which can be leveraged to directly extract information from the images~\cite{PhysRevApplied.14.014011} without having to fit the data with appropriate models.
Hence, neural networks can potentially yield an efficiency gain, as more information can be extracted per image using neural networks compared to fitting. Further, the networks can be trained to distinguish experimental noise from signal~\cite{xu2014deep}, reducing the number of pictures needed to achieve high signal-to-noise ratio.
Reducing the number of images required for a given measurement is highly important, as each cold gas realization is time consuming to prepare and is often destroyed upon imaging.

In this work, we construct a neural network model capable of extracting and processing features of one-dimensional (1d) quasi-condensate density profiles measured via absorption imaging.
We specifically train the model to estimate the temperature of the quasi-condensate directly from a single absorption image after time of flight. For such quasi-condensates, fluctuations of the density have been a key observable for the readout of many different quantities, including the temperature~\cite{PhysRevLett.96.130403, PhysRevLett.106.230405, PhysRevLett.105.230402, PhysRevA.98.043604, Langen207, Gring1318, Schweigler2021}. The fluctuations follow some underlying temperature-dependent spectrum, which can be derived theoretically. However, distortions of the fluctuations from the imaging process makes associating a given sample of fluctuations to their corresponding temperature very difficult.
By simulating the full imaging process, we can create a large number of artificial absorption images of condensates with a known temperature~\cite{schweigler2019correlations}.
Employing these images as training data for a neural network, the trained model effectively becomes a non-linear map between the measured fluctuations and the pre-imaging spectrum of fluctuations.
The paper is thus structured as follows: In section \ref{sec:theory}, we review the relevant theoretical aspects of fluctuations in one-dimensional quasi-condensates and discuss the established density ripples thermometry. Next, in section \ref{sec:setup}, we discuss the experimental setup and the architecture of the neural network. In section \ref{sec:results}, the trained neural networks are first benchmarked on simulated images and then employed on experimentally measured images. We also demonstrate how the networks extract features from the images and how these correlate with properties of the condensate. Finally, we conclude in section \ref{sec:conclusion}.

\section{Theory} \label{sec:theory}
Fluctuations play a very important role in the physics of one-dimensional many-body systems of interacting bosons. Measurements of these fluctuations, be it phase or density fluctuations, and their correlations yield important information about properties of the system~\cite{PhysRevLett.96.130403, PhysRevLett.106.230405, PhysRevLett.105.230402, PhysRevA.98.043604}. 
For instance, a well-established method of thermometry is via the observations of density ripples patterns~\cite{PhysRevA.80.033604, PhysRevA.81.031610}. These patterns emerge in the atomic density profile as the gas expands freely during time of flight before imaging. They are caused by temperature dependent, in-situ fluctuations in the phase. The temperature can be inferred from correlation functions of the density ripples, however, the measurement must be repeated many times before the correlation function can be accurately constructed. Additionally, the imaging process itself distorts the measured ripple patterns, making comparisons to theory challenging~\cite{PhysRevA.81.031610}.
Neural networks, on the other hand, do not require any knowledge of the complicated underlying theory~\footnote{In our case, knowledge of the theory is required for simulating the absorption images used for training.}. Instead, they can be trained to correlate a set of features in an image to some related property, in this case a temperature, thus making them a powerful tool for studying complex many-body systems. However, careful training of the network is necessary for achieving high-accuracy predictions. In order to have exact knowledge of the temperature of a given training example, we simulate artificial images to employ as training data.

\subsection{Fluctuations in one-dimensional condensates}
The one-dimensional (1d) Bose gas with contact interactions is described by the Lieb-Liniger model with the Hamiltonian
\begin{equation}
\begin{aligned}
\hat{H}= &-\frac{\hbar^{2}}{2 m} \int \mathrm{d} z \; \hat{\psi}^{\dagger}(z) \partial_{z}^{2} \hat{\psi}(z)+ \\
&+ \frac{g_{1 \mathrm{d}}}{2} \int \mathrm{d} z \; \hat{\psi}^{\dagger}(z) \hat{\psi}^{\dagger}(z) \hat{\psi}(z) \hat{\psi}(z) \; ,
\end{aligned}
\end{equation}
where $\hat{\psi}$ are the bosonic fields, $m$ is the mass of the atoms, and $g_{1 \mathrm{d}}$ is the interaction strength~\cite{PhysRevLett.81.938}.
Unlike their three-dimensional counterpart, one-dimensional Bose gases can not achieve long range phase order. Nevertheless, local phase coherence can be achieved in the quasi-condensate regime. In this regime the properties of the fluctuations can be described by a spectrum of Bogoliubov-type modes. The fluctuations can be expressed through the phase-density representation of the field operator~\cite{PhysRevA.67.053615}
\begin{equation}
\hat{\psi}(z)=e^{i \hat{\theta}(z)} \sqrt{n_{1 \mathrm{d}}(z)+\delta \hat{n}(z)} \; .
\end{equation}
Here, $\hat{\theta}$ is the operator describing the fluctuating phase, while $\delta \hat{n}$ are the density fluctuations relative to the mean density profile $n_{1 \mathrm{d}}$. 
Assuming a homogeneous system contained in a hard-walled box of length $L$, the fluctuations can be expanded into modes with wavenumbers $k_n = n \pi /L$ following~\footnote{eqs.~\eqref{eq:mode_density} and \eqref{eq:mode_phase} are derived assuming Neumann boundary conditions implying that the particle current vanishes on the boundary. This corresponds to the physical situation of having a hard-walled box.}  
\begin{align}
\delta \hat{n}(z)&=\sqrt{\frac{2}{L}} \sum_{n} \delta \hat{n}_{n} \cos (k_n z) \label{eq:mode_density} \\
\hat{\theta}(z)&=\sqrt{\frac{2}{L}} \sum_{n} \hat{\theta}_{n} \cos (k_n z) \label{eq:mode_phase}
\end{align}
At sufficiently large temperatures the average occupation of the observable low lying modes is much larger than one, whereby the fluctuations can be treated as classical fields $\theta (z)$ and $\delta n (z)$. Finally, in the classical field and Bogoliubov approximation, the thermal expectation values are 
\begin{align}
\left\langle\theta_{n}\right\rangle &=\left\langle\delta n_{n}\right\rangle=0 \\
\left\langle\theta_{n} \theta_{m}\right\rangle &=\delta_{n, m} \frac{k_{\mathrm{B}} T}{2} \frac{2 m}{\hbar^{2} k_{n}^{2} n_{1 \mathrm{d}}} \label{eq:var_phase} \\
\left\langle\delta n_n \delta n_{m}\right\rangle &=\delta_{n, m} \frac{k_{\mathrm{B}} T}{2}\left(\frac{g_{1 \mathrm{d}}}{2}+\frac{\hbar^{2} k_{n}^{2}}{8 m n_{1 \mathrm{d}}}\right)^{-1} \label{eq:var_density}
\end{align}
As evident of eqs.~\eqref{eq:var_phase} and \eqref{eq:var_density}, the variance of both the phase and density fluctuations scales linearly with the temperature $T$. Hence, in principle, both types of fluctuation can therefore be used for thermometry.

For thermal excitations the phase fluctuations are Gaussian, whereby the phase profile $\theta (z)$ can be described by an Ornstein-Uhlenbeck stochastic process with $z$ playing the role of time~\cite{PhysRevLett.105.015301}
\begin{equation}
    \frac{\mathrm{d}}{\mathrm{d}z} \theta(z) = f (z) \; .
\end{equation}
Here, $f(z)$ is a random force, sampled from a Gaussian distribution with zero mean, and variance $m k_{\mathrm{B}} T / \hbar^2 n_{1d}$, thus reproducing the thermal expectation values of the phase. Similarly, the thermal expectation values of the density correlations can be used as basis for Gaussian sampling of the density fluctuations.

\subsection{Formation of ripple patterns during free expansion}
In density ripples thermometry the phase fluctuations are probed through measurement of the fluctuating density pattern emerging under free expansion of the gas. The tight transverse confinement necessary for achieving a 1d gas induces a rapid transverse expansion of the gas when released, thus preventing any interactions.
Upon the free expansion, the phase gradient gives rise to a density current following
\begin{equation}
j(z)=\frac{\hbar}{m} \partial_{z} \theta(z)\left( n_{1 \mathrm{d}} + \delta n(z)\right) \; .
\end{equation}
The fluctuations of the phase field cause different regions of the gas to expand at different velocities, giving rise to the formation of density ripple patterns over time. For longer expansion times, the contrast of the density ripples increases. As discussed in Ref.~\cite{PhysRevA.80.033604}, the density ripples are distributed according to a certain power spectrum. The power spectrum is temperature dependent and can be measured experimentally, by taking the Fourier transform of the two-point density correlation function
\begin{equation}
g_{2}(\delta z)=\frac{\int \mathrm{d} z\langle n(z+\delta z) n(z)\rangle}{\int \mathrm{d} z\langle n(z+\delta z)\rangle\langle n(z)\rangle} \; .
\label{eq:g2_function}
\end{equation}
Here, the expectation value is measured experimentally by repeating the measurement several times and taking the average thereof.
Following Ref.~\cite{PhysRevA.80.033604}, $g_{2}$-functions for various temperatures can be computed and fitted to the measured results, whereby an estimate of the temperature can be made.
This process constitutes the method of density ripples thermometry.
However, the points made above also highlight some of the weaknesses of the approach, which could be overcome using neural networks.

Firstly, in order to obtain the $g_{2}$-function above, many measurements must be taken, making the method rather inefficient. This also prevents usage of the method on single images.

Secondly, the experimental imaging process itself distorts the correlation function, making a direct comparison to the theory curve difficult. 
The current solution is based on empirical studies and requires convolving simulated density profiles with an effective Gaussian point spread function (PSF) before computing the theoretical $g_{2}$-function, in order to mimic the limited resolution of the imaging system. In reality, the effective resolution of the imaging process depends on a number of parameters, including properties of the condensate itself (see Appendix~\ref{app:artificial}). Therefore, the exact width of the effective point spread function is unknown, and is hence fitted along with the temperature~\cite{PhysRevA.81.031610}.

Note, when computing the expectation value in eq.~\eqref{eq:g2_function} via averaging, it is formally assumed that all the realizations follow the same underlying power spectrum, i.e. they have the same temperature. However, small temperature variations can occur between individual realizations in an experimental setting. Thus, the temperature obtained by fitting the $g_{2}$-function is interpreted as the mean temperature of the sample.

\section{Setup} \label{sec:setup}
\begin{figure*}
\center
\includegraphics[width = 0.95\textwidth]{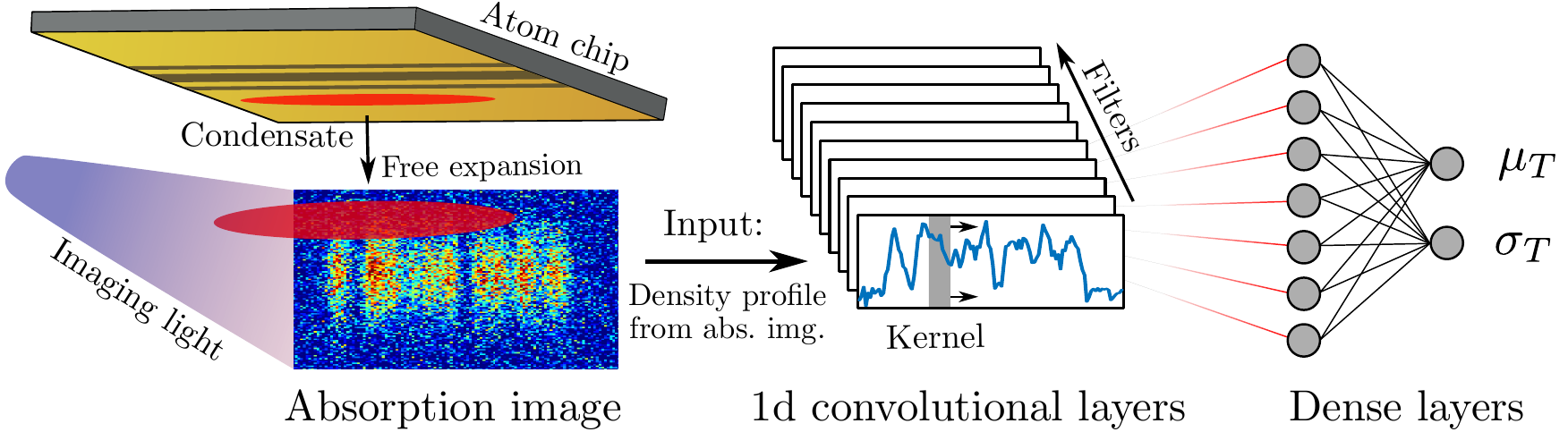}
\caption{\label{fig:network_illustration} Illustration of the application of a neural network for thermometry. The condensate is initially trapped on the atom chip trap. In order to probe it, the gas is released and undergoes free expansion, whereafter it is exposed to the imaging light. After passing through the cloud, the imaging light is captured by a camera, producing an absorption image of the cloud. By integrating out the transverse direction (here vertical) the 1d density profile is obtained and fed directly as input to the neural network.
The network consists of two stages: First, the profiles pass through a series of 1d convolutional layers, where a kernel scans over the profiles extracting important features. Next, the extracted features are passed to several dense layers, which process said features. Finally, the network outputs a normal distribution for the temperature probability, parameterized by the two output values $\mu_T$ and $\sigma_T$.
Note, layers depicted in the figure do not reflect parameters of the actual model. For the full network architecture, see table~\ref{tab:architecture}. }
\end{figure*}
When applying machine learning to an experimental setting, it is necessary to account for the whole system, including the experimental setup. Hence, there are two main approaches when it comes to training:
(i) Training the model directly on the experimental data, whereby it automatically learns how the physical system looks like perceived through the experimental measurement apparatus~\cite{rem2019identifying}, or (ii) simulating the experimental setup and thus training the model on simulated data~\cite{Picard_2019}.

The first approach is the most common, as it can circumvent any human assumptions or approximations between the neural network and the experimental interface. Especially for complex setups, it can be very challenging to theoretically describe all the experimental processes and account for all imperfections. Instead, directly connecting the model to the experimental setup circumvents the need for such approximations.

In this work we employ the alternative approach of simulating the experimental setup. In order for the model to accurately predict the temperature of a single realization, the training images must be labelled by the corresponding temperature. However, our current methods of thermometry are only capable of finding the temperature averaged over many samples. Since the temperature can fluctuate or drift throughout the experimental cycle, we have no knowledge of the temperature of the individual realizations, effectively ruling out the first approach. Additionally, simulating the absorption images is a lot quicker than producing them experimentally, and variation of key parameters can be controlled exactly. Thus, we are capable of producing large data sets necessary for a very dense sampling of the fluctuation spectra.
However, in order to simulate the training data, good knowledge of the imaging process is required. In the following we will briefly discuss the experimental setup, whereas the simulation of images is covered in the Appendix~\ref{app:artificial} and Ref.~\cite{schweigler2019correlations}.

\subsection{The experimental setup}

In the experiment, we use an atom chip trap~\cite{atomchips,folman2008microscopic,PhysRevLett.84.4749} to realize a one-dimensional quasi-condensate of $^{87}\mathrm{Rb}$ atoms. The current carrying micro-fabricated wires on the chip surface generate a very large magnetic field gradient. Combined with an external bias field from surrounding coils, the wires create a highly anisotropic Ioffe-Pritchard type trap. The trap has two tightly confined (transverse) directions with a transverse trapping frequency of $\omega_\perp = 2 \pi \times 2.1 \, \mathrm{kHz}$, and one weakly confined (longitudinal or 1d) direction. 
The magnetic trap is parabolic along the weakly confined longitudinal direction (along the main chip wire).
To achieve an arbitrary potential landscape in the longitudinal direction, we add an optical dipole potential~\cite{Tajik:19}.
The blue detuned laser light (wavelength of 660\,nm) is shaped using a DMD before reaching the magnetically trapped atoms.
In the context of this work, we employ a box trap yielding a uniform atomic density.
Thus, after evaporative cooling of the atomic gas, a homogeneous quasi-condensate of 3000-10000 atoms at a temperature down to $T = $ 20\,nK is achieved.

To measure the system, the atomic gas is released from all traps and allowed to expand freely for either 2\,ms or 11.2\,ms. Immediately thereafter, the absorption image is taken along one of the transverse directions of the cloud (parallel to the chip surface). For the imaging light we employ the $D_2$ line and use intensities around $25\%$ of saturation. The numerical aperture (NA) of our transverse imaging system is about $0.2$, and the exposure time used for the image is $75 \, \mu\mathrm{s}$. In figure~\ref{fig:network_illustration} an illustration of the imaging process and an exemplar absorption image can be found.

\subsection{The neural network model}
When setting up a neural network model, the parameterization of the problem is crucial.
The absorption images taken in the experiment yield the density of the condensate integrated over one of its transverse axis, i.e.~its 2d density. However, the density ripples exist only along the longitudinal axis of the condensate. Therefore, for the input of the network, we integrate out the transverse axis, leaving only the 1d density profile. In order to reduce the amount of photon shot-noise on the profiles, we crop the absorption images as close to the atom cloud as possible in the transverse direction.
The density profile obtained from the absorption image is then fed as input to the neural network model.

The neural network consists of two stages, as illustrated in figure~\ref{fig:network_illustration}.
The first stage of the network consists of three 1d-convolutional layers in succession, used to identify features in the density profiles.
The 1d-convolutional layers scans a kernel, which is a vector of a fixed number of weights, over the input to extract local features. Starting from one end of the input, the scalar product between the kernel and the overlapping part of the profile is taken. The resulting value is passed through a non-linear activation function, here the ReLU function $f(x) = \max (x, 0)$~\cite{nair2010rectified}. Next, the kernel moves a number of pixels specified by its stride length, and the process is repeated. As the kernel has moved across the entirety of the input profile a new profile has been produced, which effectively is the convolution between the input and the kernel. The shape of the output depends on the weights of the kernel, which through training are optimized to extract a particular feature. A single convolutional layer consists of a number of filters, each with its own associated kernel. In our model we employ 16 filters, whereby each convolutional layer is capable of detecting up to 16 unique features. 
Each convolutional layer of our model scans the output profile of the previous layer. As an imaged density profile passes through each layer, the model abstracts an increasing number of details from the profile into more general features, which in the next stage of the network are used to estimate the temperature.

The second stage of the network consist of three dense layers, which process the previously extracted features in order to produce an estimate of the temperature.
A dense layer consist of a number of neurons, where each neuron receives input from all the neurons in the previous layer. Each connection to a neuron has an associated trainable weight, which is multiplied with the input, and each neuron has a trainable bias, which is added to its output. Finally, the otherwise linear output of a neuron is passed through the activation function of the layer, making the final output non-linear. By having multiple dense layers in succession, functions of increasingly higher order can be approximated. Also for the dense layers do we employ the ReLU activation function.

Before each dense layer is placed a Dropout layer~\cite{JMLR:v15:srivastava14a}, which randomly set the output of a neuron to zero with a probability set by its dropout rate, here 0.5. This helps regularize the network to not become overly reliant on just a few neurons, which can cause overfitting. Overfitting occurs when a model learns the particular details of the training data set rather than the general features of said data. In such a case, the model will perform poorly when faced with data outside the set used for training. To check whether a model is being overfitted, it is common practice to reserve a portion of the training data as validation data, which is never shown to the model throughout training. If the model performs significantly better on the training than on the validation data, the model has become overfitted.   
Note the Dropout layers are only active while training the model. 

Following the second stage, the final layer of the network is a dense layer with two neurons, each producing an output value. The output of the model is discussed in more detail in the next section.

For additional robustness against overfitting, L2-regularization~\cite{10.5555/2986916.2987033} was employed on all kernels of the network. Furthermore, before training, each kernel was randomly initialized using the Xavier uniform scheme~\cite{pmlr-v9-glorot10a}. See table \ref{tab:architecture} for a summary of the network architecture.

\begin{table}[]
\centering
\ra{1.3}
\rowcolors{1}{}{lightgray}
\begin{tabular}{ll}
\toprule
Layer        & Parameters                                 \\ \midrule
Conv1D       & filters = 16, kernel size = 3, strides = 1 \\
Conv1D      & filters = 16, kernel size = 3, strides = 1 \\
Conv1D       & filters = 16, kernel size = 5, strides = 1 \\
Flatten      &                                            \\
Dropout      & dropout rate = 0.5                         \\
Dense        & size = 200                                 \\
Dropout      & dropout rate = 0.5                         \\
Dense        & size = 200                                 \\
Dropout     & dropout rate = 0.5                         \\
Dense        & size = 200                                 \\ 
Dense (output)       & size = 2                         \\ \bottomrule
\end{tabular}
\caption{\label{tab:architecture} Neural network architecture. The activation function of each layer is the ReLU function~\cite{nair2010rectified}, except for the output layer which employs the softplus function. Each kernel is initialized using the Xavier uniform scheme~\cite{pmlr-v9-glorot10a} and uses L2 regularization~\cite{10.5555/2986916.2987033}.}
\end{table}

\subsection{Training the model}
The behaviour of a given model is learnt throughout its training, where the different weights and biases of the network are optimized to minimize some function (loss) of the output of the network when applied to the training data. We employ supervised learning, where the network is trained to approximate a function that maps an input to an output based on examples of input-output pairs, here the density profiles and associated temperatures.

Due to the stochastic nature of the fluctuations in the condensate, exactly determining the temperature from a single image is practically impossible, as the fluctuations of a single realization only sample a small portion of the underlying fluctuation spectrum. Therefore, we have designed and trained the network to also return an uncertainty of the temperature estimate. This is achieved by outputting a normal distribution for the temperature probability of a given condensate, parameterized by two values $\mu_T$ and $\sigma_T$, denoting the mean and standard deviation, respectively. Since both parameters are positive, we employ the softplus activation function $f(x) = \log ( 1 + e^x )$ for the output layer with two neurons~\footnote{The softplus function ensures that the outputs $\mu_T$ and $\sigma_T$ are positive. The softplus function has essentially the same effect as the ReLU function, however, it is more numerically stable when employing logarithms in the loss function.}.
To attain the desired behavior, we trained our model using the negative-log-likelihood function of our output distribution as the loss
\begin{equation}
    \mathrm{loss} (T) = \log \left( \sigma_T \right) + \frac{1}{2 \sigma_T^2} \left( T - \mu_T \right)^2 \; ,
    \label{eq:loss}
\end{equation}
where $T$ is the target temperature. The likelihood function maximizes the overlap between the output distribution and the target, while the logarithm negates the possibility of numerical underflow~\cite{Goodfellow-et-al-2016}.

At the end of training, $\sigma_T$ can be considered an uncertainty estimate of the models own temperature prediction, which is evident from eq.~\eqref{eq:loss}: If the model is uncertain of the temperature, it will output a wide distribution (large $\sigma_T$) to minimize the contribution of the second term in eq.~\eqref{eq:loss} scaling as the error of the estimate squared. Conversely, if the model is certain of its prediction, the error is likely small, whereby the minimum loss is obtained for a narrow distribution (small $\sigma_T$ mimizing the first term of eq.~\eqref{eq:loss}).
Even for a perfect model, $\sigma_T$ won't be vanishing, as the inherent randomness of the fluctuations combined with the sparse sampling makes an exact temperature prediction impossible.

The choice of loss function determines the behaviour of the model. By changing the loss function and pairing each density profile with a different parameter, one can easily train a new model using the same network architecture to estimate a different property of the quasi-condensate.

We wish the neural network models to be applicable for practically any quasi-condensate we can experimentally realize in a box-trap.
Hence, we have trained the models on simulated images of condensates in box-traps of lengths 50\,$\mu\mathrm{m}$ to 150\,$\mu\mathrm{m}$ and with atomnumbers of 4000 to 10000. Condensates realized in the experiment are typically in the range 20\,nK to 90\,nK, however, we have also trained the models on higher temperatures (up to 160\,nK) in an effort to reduce potential biases in their predictions. The imaging intensity was set to $25\%$ of saturation.
We trained two separate models for short (2\,ms) and long (11.2\,ms) expansion time on batches of 64 profiles at a time using the Adam optimizer~\cite{kingma2017adam} with a learning rate of 0.001. After 100 epochs (cycles of full training set) we observed consistent convergence of the loss. The loss for the training and validation data (10$\%$ of the training data chosen randomly at the start of training) would converge to the same value, showing no signs of overfitting.
For more details of the training data set and discussion of the training strategy employed see Appendix~\ref{app:training_strategy}. 

\subsection{Combining multiple predictions}
The accuracy of the temperature prediction from a single density profile is physically limited by the stochastic nature of the quasi-condensate. However, a more accurate estimate of the temperature can be obtained by combining predictions on individual profiles.
Given a set of $N$ density profiles, an estimate of the mean temperature is simply the mean of the individual predictions $\mu_T^{(i)}$
\begin{equation}
    \bar{\mu}_T = \frac{1}{N} \sum_{i=1}^N \mu_T^{(i)}  \; .
    \label{eq:mean_estimate}
\end{equation}

Next, we employ the standard error of the mean as the uncertainty of $\bar{\mu}_T$. Given a set of $N$ density profiles, we assume the neural network to produce $N$ independent normal distributions for the probability of the condensate temperature. Combining the means and variances of said $N$ distributions, we obtain the following expression for the uncertainty
\begin{equation}
    \Delta_T = \frac{1}{N} \sqrt{ \sum_{i=1}^N \left( \mu_T^{(i)} - \bar{\mu}_T \right)^2 + \sum_{i=1}^N ( \sigma_T^{(i)} )^2  } \; .
    \label{eq:uncert_estimate}
\end{equation}
Importantly, eq.~\eqref{eq:uncert_estimate} depends both on the variance of the individual predictions $\mu_T^{(i)}$ around the mean $\bar{\mu}_T$ and on the individual variances $(\sigma_T^{(i)})^2$.

Considering a normal probability distribution output by the neural network, one could randomly sample said distribution a large number of times. Given $N$ different normal distributions, sampling each distribution and taking the mean of all the samples will reproduce the mean $\bar{\mu}_T$ of eq.~\eqref{eq:mean_estimate}, while computing the standard error of the mean of all the samples will indeed yield $\Delta_T$ of eq.~\eqref{eq:uncert_estimate}.

\section{Results} \label{sec:results}
In this section we evaluate the performance of our trained networks. First, we will benchmark the models on simulated data, where the temperatures of the different density profiles are already known. Next, we apply our model to experimental data and compare its predictions to the results of density ripples thermometry. Finally, we study the output of the different convolutional layers of the network in order to gain insight into which features are extracted by the model and how they relate to the underlying physics of the quasi-condensate.

\subsection{Benchmarking on simulated data}
\begin{figure}
\center
\includegraphics[width = 0.95\columnwidth]{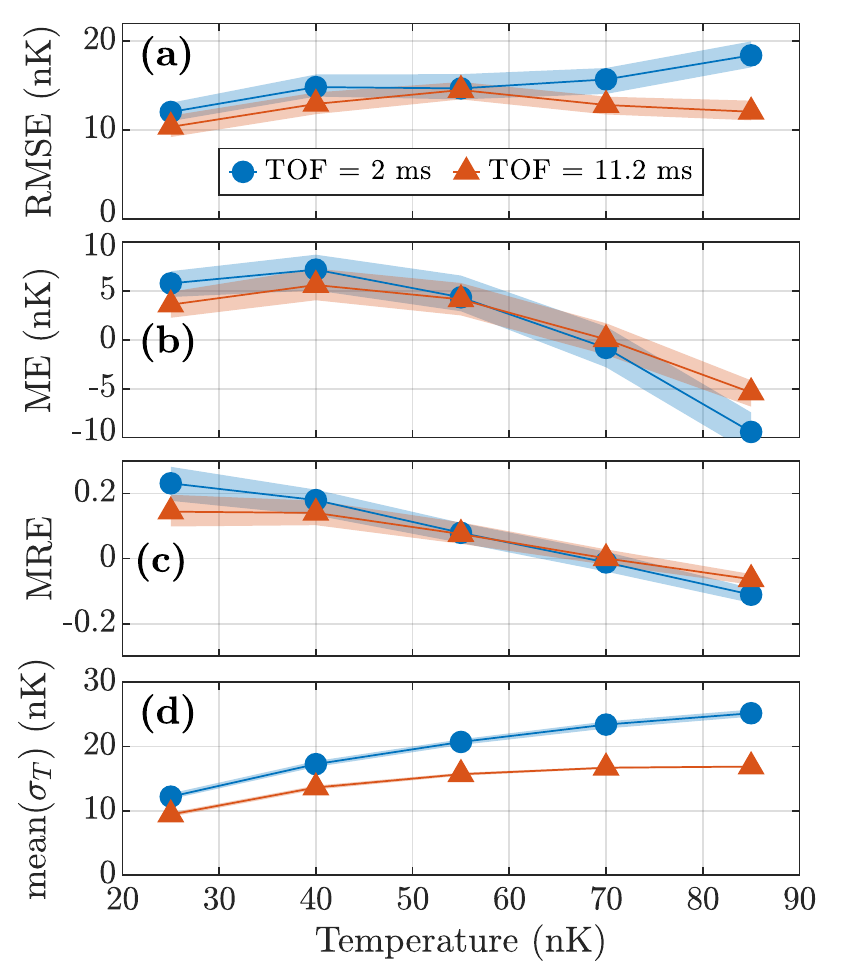}
\phantomsubfloat{\label{fig:training_a}}
\phantomsubfloat{\label{fig:training_b}}
\phantomsubfloat{\label{fig:training_c}}
\phantomsubfloat{\label{fig:training_d}}

\vspace{-3\baselineskip}
\caption{\label{fig:training} Performance of neural network models on simulated data for 2\,ms and 11.2\,ms expansion time (TOF).
(a) Root-mean-square error (RMSE, eq.~\eqref{eq:RMSE}), (b) mean error (ME, eq.~\eqref{eq:ME}), and (c) mean relative error (MRE, eq.~\eqref{eq:MRE}) of the neural networks individual predictions $\mu_T$. (d) Mean value of the uncertainties of the predictions $\sigma_T$. For each temperature, the averages were computed over predictions on $M=250$ randomly generated profiles. The shaded areas show 95$\%$ confidence intervals obtained via bootstrapping (bias corrected accelerated method~\cite{10.1214/ss/1032280214}) with $M$ re-samplings.}
\end{figure}

For the benchmark we generate new test data in order to discern whether the trained neural network models are capable of \textit{generalization}, i.e. making sensible predictions for data entirely separate from the training data. 
Here we will focus on predictions on single profiles, while benchmarks of combined predictions on multiple profiles can be found in Appendix~\ref{app:benchmark}. 

We will be benchmarking the performance on temperatures ranging from 25\,nK to 85\,nK in steps of 15\,nK, thus sampling the typically achieved temperatures in the experiment. 
For each temperature, as well as for short (2\,ms) and long (11.2\,ms) expansion time (TOF), we simulate $M = 250$ density profiles.
For each profile, the number of atoms is randomly chosen as either 4000, 7000 or 10000. 
Similarly, the length of the box-trap is either 50, 100 or 150\,$\mu\mathrm{m}$. 
After simulating the profiles, each single profile is fed to the neural network and a prediction  of the temperature is obtained in the form of the normal distribution parameterized by $\mu_T$ and $\sigma_T$.
Then we quantify the accuracy of the networks predictions by computing the root-mean-square error (RMSE)
\begin{equation}
    \mathrm{RMSE} \, (T) = \left( \frac{1}{M} \sum_{i= 1}^M \left(\mu_T^{(i)} - T \right)^{2} \right) ^{1/2} \; ,
    \label{eq:RMSE}
\end{equation}
the mean error (ME)
\begin{equation}
    \mathrm{ME} \, (T) = \frac{1}{M} \sum_{i= 1}^M \left( \mu_T^{(i)} - T \right) \; ,
    \label{eq:ME}
\end{equation}
and the mean relative error (MRE)
\begin{equation}
    \mathrm{MRE} \, (T) = \frac{1}{M} \sum_{i= 1}^M \frac{ \mu_T^{(i)} - T }{T} \; ,
    \label{eq:MRE}
\end{equation}
which we have plotted in figures \ref{fig:training_a}, \ref{fig:training_b}, and \ref{fig:training_c}, respectively.

Starting with figure \ref{fig:training_a}, we observe an RMSE between 10\,nK and 15\,nK, with the exception of the predictions on condensates of $T =$ 85\,nK for short expansion time, where the RMSE is slightly larger, around 19\,nK.
The RMSE does not reveal whether there is any systematic errors, or bias, in the predictions of the models. While the loss function~\eqref{eq:loss} is symmetric with respect to temperature, the physical problem is not, as temperature differences become increasingly harder to discern at larger temperatures. Therefore, during training, the models are susceptible to becoming biased. For additional discussion of the bias and possible strategies to overcome it, see Appendix~\ref{app:training_strategy}.
To check for potential biases, we plot in figures~\ref{fig:training_b} and \ref{fig:training_c} the mean error and mean relative error, respectively. The observed non-zero mean error indicates the presence of a bias in our models, albeit a relatively small one. Particularly the model trained on long expansion times has an absolute mean error of less than 5\,nK, corresponding to a mean relative error of about 15\% for the lowest sampled temperature. 

Finally, in figure \ref{fig:training_d} we plot the mean value of the standard deviations output by the network $\sigma_T^{(i)}$~\footnote{We also compute confidence intervals using bootstrapping for the mean of $\sigma_T^{(i)}$, however, the interval is so small that it is hardly visible in the plot.}. For all temperatures benchmarked on, the mean value of $\sigma_T$ is similar to, or greater than, the RMSE, confirming that the models prediction $\sigma_T$ can reliably be interpreted as an estimate of their own uncertainty.
Further, we find the uncertainty to increase as the temperature rises, which is consistent with the linear temperature dependence of the variance of the thermal expectation value of the condensate fluctuations, eqs.~\eqref{eq:var_phase} and \eqref{eq:var_density}. 
Lastly, the uncertainty predictions of the model trained at 2\,ms expansion time are noticeably larger than those of the model trained for 11.2\,ms expansion. At larger expansion times the contrast of the density ripples patterns is much higher, making it much easier to distinguish patterns for different temperatures. Thus, the greater uncertainties seen for 2\,ms expansion time are expected.

\subsection{Application on experimental data}

Having benchmarked the models, we are now ready to apply them to experimental data and compare their performance with the established density ripples thermometry~\cite{PhysRevA.81.031610}. 
We consider a measurement of a condensate of 10000 atoms trapped in a 100\,$\mu\mathrm{m}$ box-trap followed by 11.2\,ms of free expansion. For the imaging, an intensity of 24(2)$\%$ saturation was used. The measurement was repeated 180 times, each requiring a separate realization of the condensate.

\begin{figure}
\center
\includegraphics[width = 0.95\columnwidth]{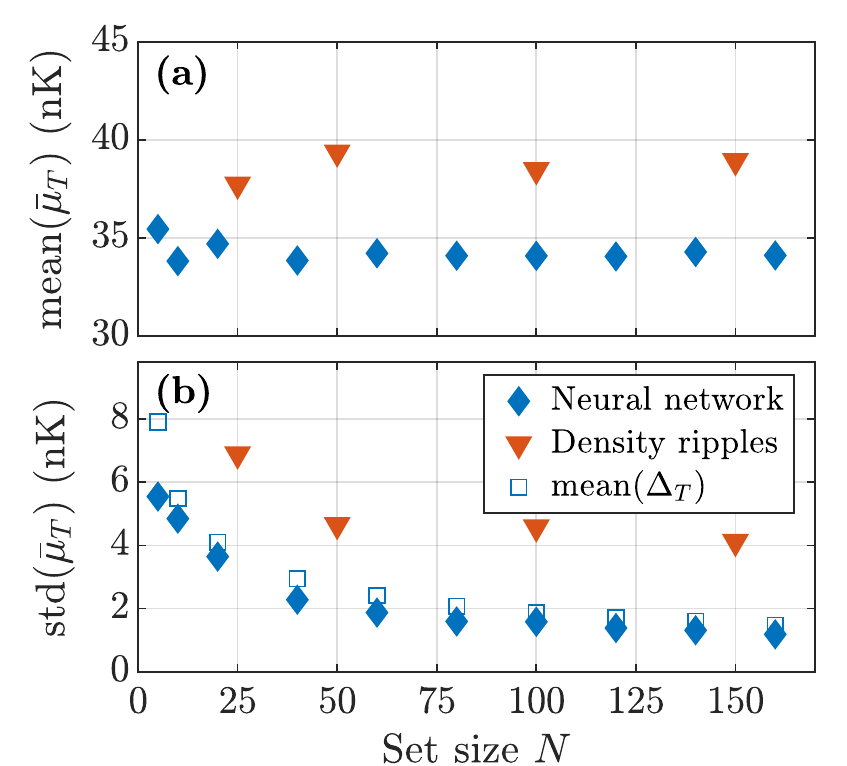}
\phantomsubfloat{\label{fig:thermometry_TOF_a}}
\phantomsubfloat{\label{fig:thermometry_TOF_b}}

\vspace{-1\baselineskip}
\caption{\label{fig:thermometry_TOF} Benchmark of thermometry with neural network model and density ripples method on experimental data taken at 11.2\,ms expansion time.
The two methods are employed on subsets of size $N$ sampled from the full set of measured density profiles using bootstrap techniques (see main text).
(a) Predicted mean temperatures $\bar{\mu}_T$ averaged over the bootstrap samples for various $N$.
(b) Standard deviation of $\bar{\mu}_T$ over the bootstrap samples for various $N$. Additionally, the uncertainty of the neural network $\Delta_T$ averaged over the bootstrap samples is plotted.}
\end{figure}

\begin{figure}
\center
\includegraphics[width = 0.95\columnwidth]{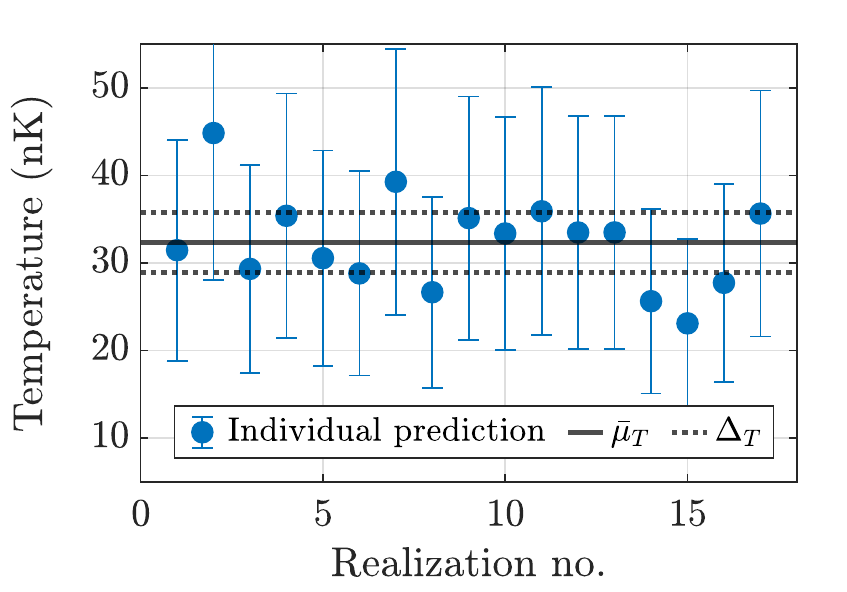}

\caption{\label{fig:thermometry_insitu} Application of neural network model on experimentally measured profiles after 2\,ms expansion time. The data was taken immediately prior to the data presented in fig. \ref{fig:thermometry_TOF}. The points indicate the temperature predictions for the individual profiles $\mu_T^{(i)}$, while the error bars denote the individual uncertainties $\sigma_T^{(i)}$. The solid line marks the mean of the individual predictions $\bar{\mu}_T = 32.4\,\mathrm{nK}$, and the dashed lines mark the uncertainty of the mean $\Delta_T = 3.4\,\mathrm{nK}$.}
\end{figure}

We proceed to test the accuracy of the two thermometry methods at various numbers of measurement repetitions by performing bootstrap sampling~\cite{EfroTibs93} of the full measurement set:
First, a subset of $N$ profiles is randomly drawn, with replacement, from the full set of 180 measured density profiles. For the given subset, its mean temperature $\bar{\mu}_T$ and the uncertainty thereof are estimated using the two methods~\footnote{For brevity, we employ the notation $\bar{\mu}_T$ for mean temperature estimates obtained via both the neural network and the density ripples thermometry.}. The process of sampling and estimating is repeated $D=50$ times, producing a set of bootstrap sampled mean temperature predictions $\{ \bar{\mu}_T^{(i)} \}_{i = 1}^D$ for each of the two methods (and a set $\{ \Delta_T^{(i)} \}_{i = 1}^D$ for the neural network).
Finally, the bootstrap sampling is performed for a number of different set sizes $N$~\footnote{The density ripples thermometry is applied to only a few set sizes $N$, as it is substantially slower than the neural network.}.

In figure~\ref{fig:thermometry_TOF_a} we plot, for each $N$, the averages of the mean temperature bootstrap sets $\{ \bar{\mu}_T^{(i)} \}_{i = 1}^D$. For smaller set sizes $N$, the sampled subsets might not be fully representative of the full population, whereby we observe slight variations in the average temperatures. However, as $N$ grows sufficiently large, the predictions of both the neural network and the density ripples thermometry settle on fixed temperatures.
Interestingly, the two methods settle on two slightly different mean temperatures, with the density ripples thermometry converging to a temperature of around 39\,nK, while the neural network predicts a mean temperature of 34\,nK.
In a more detailed study of the density ripples thermometry (see Ref.~\cite{schweigler2019correlations}), the method was applied to simulated data for a wide range of temperatures. The predictions of the density ripples method were shown on average to be around 10$\%$ higher across all temperatures, which is close to the difference in temperature prediction seen in figure~\ref{fig:thermometry_TOF_a}.

Next, we compare the precision of the two methods by taking the standard deviation of the mean temperature bootstrap sets $\{ \bar{\mu}_T^{(i)} \}_{i = 1}^D$. 
The results are plotted in figure~\ref{fig:thermometry_TOF_b}, where we observe the neural network exhibiting a much smaller variation in its predictions compared to the density ripples thermometry.
Indeed, the same precision of the density ripples thermometry applied to $N=100$ experimentally measured density profiles (which is a typical number of repetitions for this type of measurement~\cite{schweigler2019correlations}) can be achieved by the neural network using less than a quarter of the profiles.

By virtue of the bootstrap sampling, the standard deviations plotted in figure~\ref{fig:thermometry_TOF_b} can be considered estimators of the uncertainties of the two methods~\cite{EfroTibs93}. 
However, the neural network is also capable of predicting its own uncertainty $\Delta_T$, which can readily be computed via eq.~\eqref{eq:uncert_estimate}. Thus, for each set size $N$, we can average the bootstrap samples of the uncertainty $\{ \Delta_T^{(i)} \}_{i = 1}^D$ and plot them for comparison in figure~\ref{fig:thermometry_TOF_b}. For small $N$, we observe some discrepancy between the averaged uncertainty  and the standard deviation of the mean temperature predictions. However, we start to observe good agreement between the two as the bootstrap sampled subsets become increasingly representative of the population (as $N$ increases).
Thus, the comparison demonstrates that the measure of uncertainty $\Delta_T$ given by eq.~\eqref{eq:uncert_estimate}, which combines the individual predictions of the neural network on all profiles contained in a given set, correctly describes the uncertainty of the prediction $\bar{\mu}_T$.\\

Next, we turn to the thermometry of the condensate after only 2\,ms expansion time, where the measured longitudinal density profile highly resembles its in-trap shape. Hence, the contrast of the density ripples pattern is very low, and the density ripples thermometry does not apply.
Having temperature estimates for individual or small sets of profiles at short expansion time has recently become very relevant in our experimental setup, as such profiles are used for real-time optimization of an optical dipole trap projected unto the atoms via a DMD~\cite{Tajik:19}. 
We consider data taken immediately prior to the one presented in figure~\ref{fig:thermometry_TOF}. Therefore, the temperature of the quasi-condensate should be very similar here. The intensity of the imaging light shone on the cloud was around 23(2)$\%$ of saturation. Only 17 images were taken for this measurement.

The thermometry results are plotted in figure~\ref{fig:thermometry_insitu}. We find the predictions of the neural network model to exhibit rather low variance, with the exception of a few outliers, thus demonstrating consistency and reliability.
Further, variations in the condensate temperature can occur between different experimental realizations, as experiment drifts and other imperfections can influence it. Indeed, we find a few consecutive outliers (realizations 14-16) with predicted temperature lower than the mean, which could indicate an experimental drift.
Since no other methods of thermometry exist in this regime for such a small number of repetitions, the only measure of accuracy of the models predictions is by comparison with the results for long expansion time, which were measured immediately after. Indeed, we find the two models trained on separate data sets of different expansion time to predict very similar mean temperatures (32.4\,nK for 2\,ms TOF and 34\,nK for 11.2\,ms TOF), thereby further increasing our trust in the accuracy and reliability of the method.

\begin{figure*}
\center
\includegraphics[width = 1\textwidth]{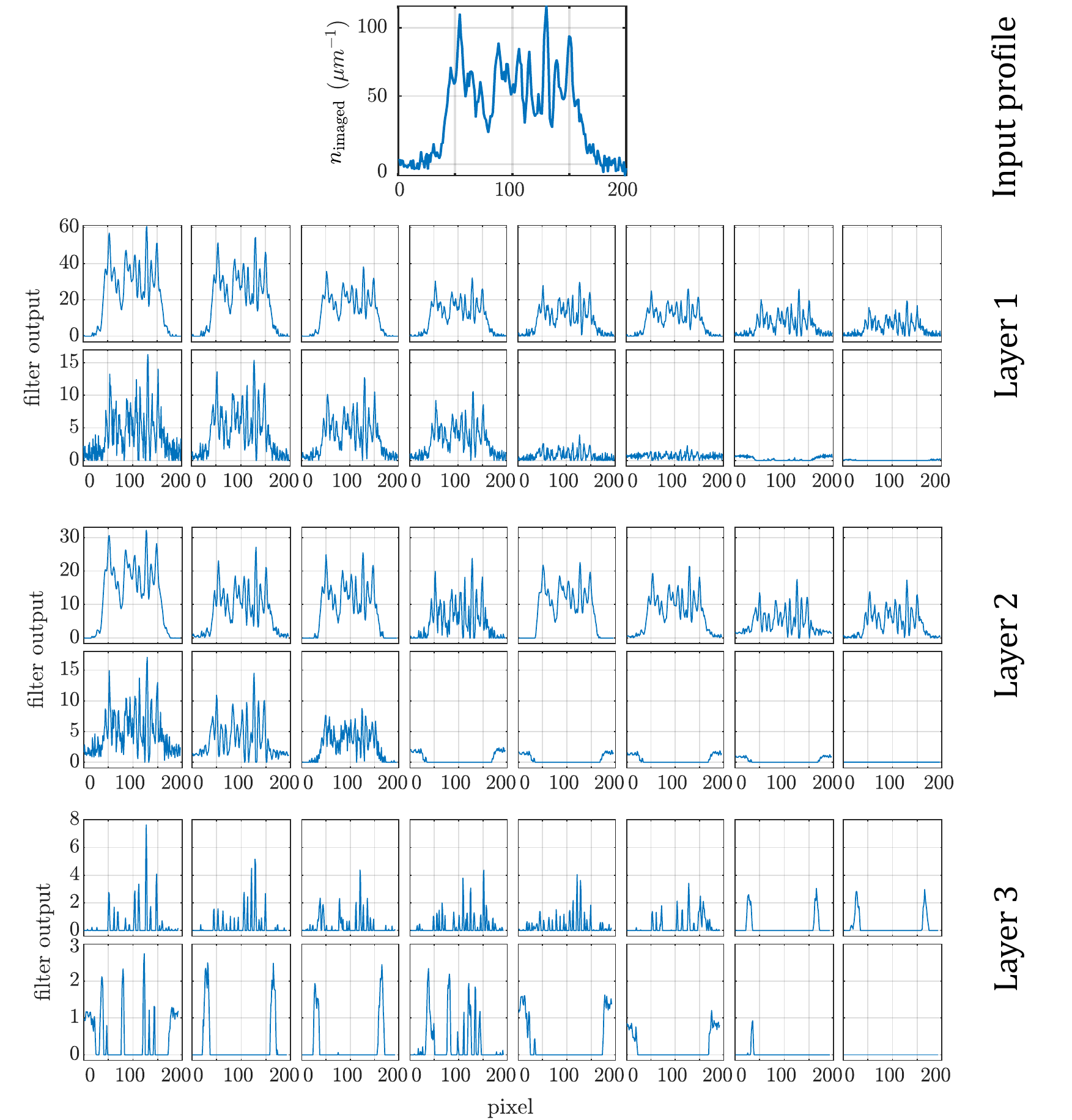}

\caption{\label{fig:feature_maps} Visualization of feature maps for the first stage of the neural network, consisting of three 1d-convolutional layers each with 16 filters. First, the imaged density profile is passed as input to the first convolutional layer, where the outputs of its filters are sorted by max value and plotted. Next, all the outputs of the first layer are passed as input to the second layer, where each filter scans all the inputs (order of inputs/filters does not matter) and combines them in a single output profile. Again the output of each filter is plotted. The process is repeated as the outputs of the second convolutional layer are passed as input to the third.
As a density profile is propagated deeper within the network, an increasing number of general features are extracted from it.
The outputs of the third layer are passed to the second stage of the network for processing.}
\end{figure*}

\begin{figure*}
\center
\includegraphics[width = 1\textwidth]{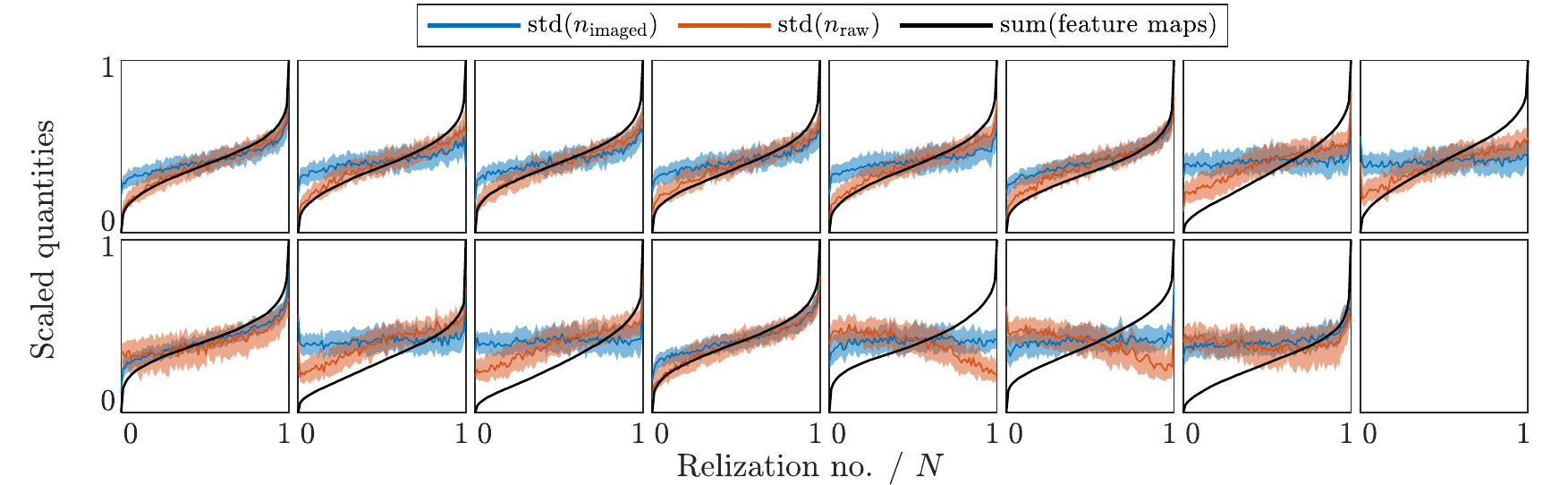}

\caption{\label{fig:physical_quant} Comparison of physical properties of atomic density profiles with the corresponding outputs of the third convolutional layer of the neural network.
The subplots represent the 16 filters of the layer and are plotted in the same order as in figure~\ref{fig:feature_maps}.
Starting with $N = 38000$ individual simulations of quasi-condensates at different temperatures, the imaged profiles were passed through the network, and the output of each filter was summed. For comparison, the standard deviation of the atomic density profile both before and after simulating the imaging process was taken.
For each filter, the results were sorted according the filter output and scaled to the interval [0, 1].
A moving median along with 25- and 75-percentiles of the standard deviations were taken over a window of 100 realizations, and they are plotted as a solid line and shaded area, respectively.
The final panel has been left blank, as the output of the corresponding filter is zero.
}
\end{figure*}

\subsection{Visualizing feature maps}

In the previous two subsections we demonstrated the predictive power of our neural network models, however, it is still unclear how exactly those models arrive at their predictions. Neural networks are generally opaque, as their predictions are based on the specific values of their weights and biases obtained through training.
However, by studying the outputs of the 1d-convolutional layers in the first stage of the network, we might be able to understand which features of the quasi-condensate density profiles are extracted and used for the temperature prediction.

Directly examining the learnt weights of the filters kernels does not reveal much information regarding the actual features extracted. Instead, we pass a single profile through the first stage of the network and plot the output of each layer in figure~\ref{fig:feature_maps}, creating what is known as a feature map.
The input density profile is simulated for a quasi-condensate of 80\,nK having undergone 11.2\,ms free expansion.

Looking at the outputs of the individual filters of the first layer, many features of the input profile are still clearly visible. The filters plotted in the second row appear to target noise on the profiles, with two filters of the first layer capable of separating the density profile from the outer regions of the image containing only photonic shot-noise. Note, no output values of the filters are negative, due to the ReLU activation function.

Similarly, one can still recognize features of the initial profile in some of the outputs of the second layer. However, several filters now completely exclude the region with atoms and only output the background noise. Given the number of similar filters, and one filter having basically no output at all, there clearly is some redundancy in the network. While too much redundancy is inefficient, having a few filters extract similar features can protect against overfitting. 

In the third and final layer, most features of the input profile have been abstracted away. The filters in this layer appear to specialize in one (or more) of the following three areas: Local density variations, background noise, and edge detection.
A number of filters, in particular the first six, output several well-separated spikes all located within the region of the quasi-condensate. The location of the spikes appears correlated to the density ripples, but the outputs do not directly resemble the input profile to the same degree as outputs in previous layers. These layers likely output the local density variance or some derivative quantity thereof.
Several other filters of the third layer output just two, roughly symmetric spikes. We believe these filters to be responsible for edge detection, i.e. indicating the border between the atomic cloud and the background. 
Lastly, a few filters output the isolated photon shot-noise from the background. Knowing the level and signature of the background noise is important, as the same type of noise will be hidden among the density fluctuation of the condensate. Hence, the network factors the output of these filters into its prediction of its own uncertainty.

Finally, we wish to quantitatively study the relation between physical properties of the quasi-condensate and the outputs of the first stage of the network. As discussed in Section~\ref{sec:theory}, the in-trap density fluctuations and the density ripples, which both are variances in the atomic density profile, depend on the temperature. Therefore, one of the most likely properties extracted from the profiles is the standard deviation of the density.
Hence, we simulate 38000 density profiles at various temperatures for 11.2\,ms expansion time and pass them through the first stage of our network. Although the output of the third convolutional layer is rather abstract, we suspect it to be related to the local density variance, whereby we employ the sum of the output from each of its filters as our quantitative measure.
For comparison we select the part of the profiles containing the atoms and compute the standard deviation, both before and after simulating the imaging process. 
Since our interest mainly concerns the overall trend, we re-scale all the results to the interval [0, 1] and sort them according to the output of the neural network.
Due to the stochastic nature of the quasi-condensate and the large number of samples, we find quite a few statistical outliers in the standard deviations of the density profiles. To clean up the results, we therefore compute the moving median and the 25- and 75-percentiles of the standard deviations over a window of 100 realizations. Note, the moving median/percentiles are taken over the sorted data, and the window size has negligible influence on the results.

The comparison between the standard deviations of the density profiles and the outputs of the first stage of the neural network is plotted in figure~\ref{fig:physical_quant}. Each tile contains the output from one filter, and the tiles are plotted in the same order as in figure~\ref{fig:feature_maps}. Comparing the figures~\ref{fig:feature_maps} and~\ref{fig:physical_quant}, we observe a remarkably good agreement between the outputs of the filters concerned with density variations (in particular the first six filters) and the standard deviation of the atomic density profiles before imaging. Further, the standard deviations taken before and after imaging clearly differ, with the network consistently matching the one pre-imaging.
Importantly, the finite resolution of the imaging process, the movement and blurring of the atomic cloud during exposure, and the addition of photon shot-noise substantially alter the imaged density profiles. Hence, these effects have to be accounted for in contemporary methods, such as the density ripples thermometry fitting an effective PSF to the data. In other words, simply taking the variance of the profiles after imaging will not yield a good estimate of the temperature. The neural network model has no knowledge of the imaging process but is instead trained to approximate a map between the imaged profiles and their temperatures.
The results of figure~\ref{fig:physical_quant} should not be interpreted as the network undoing the effect of the imaging, but rather as the standard deviation of the density profile pre-imaging being the most prominent physical feature dependent on the temperature, of which the network maps to.
It is also worth noting that the filters responsible for edge detection and background noise show only little or no correlation to the density variations.

Thus, by analysing the first stage of the network using feature maps we have shown a high correlation between the output of the full network (the temperature estimate and its uncertainty) and local variations in the atomic density profile before imaging.
From our preexisting knowledge of the theory of quasi-condensates, we already knew that such local density variations are highly temperature dependent. However, the same type of analysis can be extended to lesser known systems, enabling possible discoveries of physical correlations and dependencies.

\section{Conclusion} \label{sec:conclusion}
We have designed and trained neural network models to predict the temperature of individual realizations of one-dimensional Bose gases in the quasi-condensate regime. For each input density profile, the models output a normal distribution parameterized by its mean and standard deviation for the probability of the temperature. We use the mean as the temperature estimate and interpret the standard deviation as the models estimate of its own uncertainty. We trained two separate models for short (2\,ms) and long (11.2\,ms) expansion time on simulated data.

First, the models were benchmarked on simulated data with known temperature, revealing the models to be rather accurate, as the root-mean-square-error of their temperature predictions on individual predictions was mostly between 10-15\,nK for the temperature range achievable in the experiment (20\,nK to 90\,nK). Further, the predicted standard deviations $\sigma_T$ were similar to the root-mean-square-errors, demonstrating that they reliably can be interpreted as uncertainties of the models temperature predictions.

For thermometry on experimental data taken with long (11.2\,ms) expansion time of the condensate, we employed bootstrap sampling techniques to compare the performance of the neural network model to that of the established density ripples thermometry when applied on data sets of various sizes.
The two methods exhibited a slight discrepancy in predicted mean temperature when averaged over the bootstrap samples, with the neural network predicting a mean temperature around 34\,nK, while the density ripples thermometry predicted a mean temperature of 39\,nK. However, the density ripples thermometry has been demonstrated to be slightly positively biased, which could explain the difference.
The precision of the methods was compared by taking the standard deviation of their mean temperature predictions over the bootstrap samples. Here, the neural network model exhibited much smaller variation in its predictions, needing only a fraction of the number of density profiles to achieve similar precision as the density ripples thermometry.
Further, we compared the standard deviation of the neural networks mean temperature predictions, which is an estimator of the methods uncertainty, to its own predicted uncertainty averaged over the bootstrap samples. There we observed excellent agreement, thus demonstrating the neural networks ability to accurately predict its own uncertainty.

For a sample of density profiles measured at short (2\,ms) expansion time, the temperature predictions of the neural network model exhibit a relatively small variance with most of them close to the mean temperature of $32.4 \pm 3.4$\,nK. The measurements with short and long expansion time were taken in succession, whereby the very similar mean temperature predictions of the two separate models further enforce our confidence in the method. 

Lastly, we have studied the extraction of features from the density profiles occurring in the convolutional layers of the model by constructing a number of feature maps. By comparing the feature maps with properties of the density profiles, we find that the model is capable of inferring the summed density variations of the profiles pre-imaging. This result shows great promise for the application of neural networks in finding correlations between different physical parameters of a system. Importantly, by modifying the loss function and pairing the images with a different parameter, the neural network can easily be re-trained to extract different properties of the system.

\section*{Acknowledgements}
We thank S{\o}ren Meldgaard for enlightening discussions. F.M., J. Sabino, and F.C. acknowledge the support of the Doctoral Program CoQuS. T.S. acknowledges support from the Max Kade foundation through a postdoctoral fellowship. J. Sabino thanks the support from Funda\c{c}\~{a}o para a Ci\^{e}ncia e a Tecnologia (Portugal), namely through project UIDB/EEA/50008/2020. J. Sabino also acknowledges the support from the DP-PMI and FCT (Portugal). This research was supported by the ESQ (Erwin Schr\"{o}dinger Center for Quantum Science and Technology) Discovery programme, hosted by the Austrian Academy of Sciences (\"{O}AW), the SFB 1225 ’ISOQUANT’ (grant number I3010-N27), financed by the Austrian Science Fund (FWF), and the Wiener Wissenschafts- und TechnologieFonds (WWTF) project No MA16-066 (SEQUEX).

\appendix

\section{Training strategy} \label{app:training_strategy}
\begin{figure*}
\center
\includegraphics[width =0.9\textwidth]{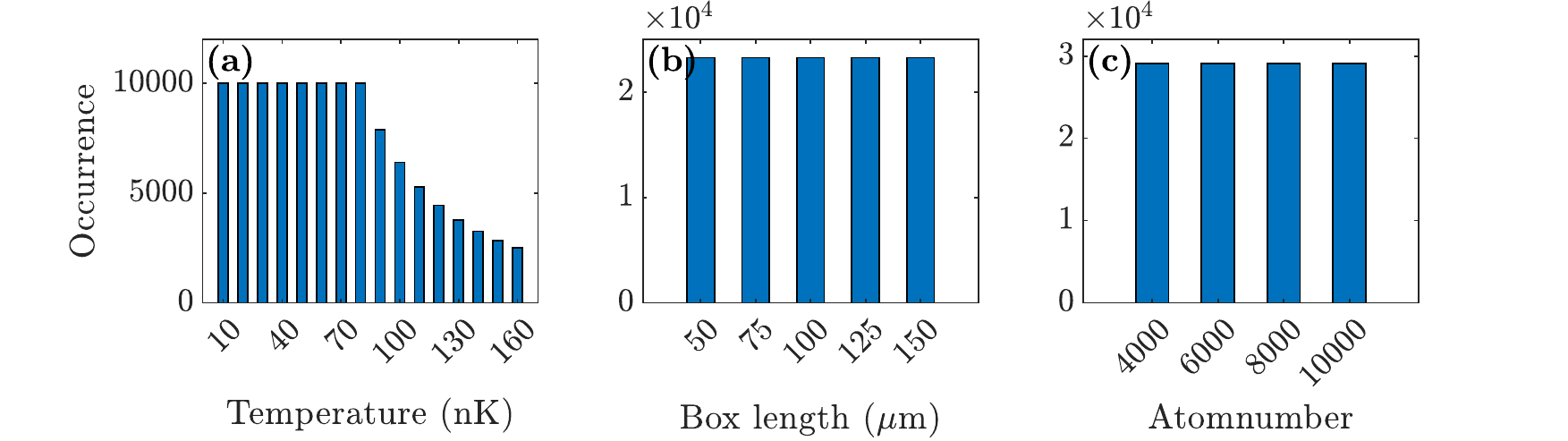}
\phantomsubfloat{\label{fig:training_data_a}}
\phantomsubfloat{\label{fig:training_data_b}}
\phantomsubfloat{\label{fig:training_data_c}}

\vspace{-1\baselineskip}

\caption{\label{fig:training_data} Distribution of condensate parameters for the simulated density profiles used for training of the neural networks. (a) Distribution of temperatures. For temperatures larger than the ones typically found in the experiment, the occurrence of said temperatures decreases to proportionally reduce their contribution to the loss function during training. This results in an a smaller bias of the model for the target temperatures. (b,c) Distribution of box (condensate) lengths and atomnumbers, respectively. Unlike the temperature, both of these parameters are distributed uniformly throughout the training data set.}
\end{figure*}

One of the main challenges of using neural networks for thermometry of quasi-condensates is training unbiased models, i.e.\ models whose temperature predictions are distributed evenly around the true temperature.
As detailed earlier, we employ supervised learning to train our models. The models are fed a batch of profiles, and a temperature prediction is made for each single profile. Then the loss function, here the negative log-likelihood function~\eqref{eq:loss}, is computed using the known, true temperatures of the profiles. Finally, the average value of the loss function over the batch of profiles is computed. By updating the weights and biases of the neural network in order to minimize the average loss, the model attains the desired behavior. 

While the loss function itself is unbiased (the mean-squared-error is symmetric around the true temperature $T$), the physical problem is not. For higher temperatures it becomes increasingly difficult to discern two condensates of similar temperature, whereby the uncertainty $\sigma_T$ becomes temperature dependent. Further, the training set only extends to a certain, maximum temperature. Thus, during training the network will never be exposed to temperatures beyond said extremum and is therefore highly unlikely to predict temperatures beyond the range on which it is trained, hence introducing additional bias.

Our aim is to train neural networks capable of accurately estimating the temperature of a single condensate realized in our experiment. Since our experiment produces condensates in the range 20\,nK to 90\,nK, we are willing to sacrifice performance at all other temperatures if it results in a higher accuracy within the range of interest. 
In order to reduce the bias of the model, two main options are available: Either the loss function should be modified to account for the temperature dependent uncertainty, or the training data set should be selected appropriately. 

Changing the loss function is problematic, as the uncertainty itself is an output of the neural network, whereby its dependence on temperature is not known until the network has been trained and benchmarked. If we change the loss to account for one dependence and then re-train the network, the temperature dependence of the uncertainty might have changed. 
Instead, we have selected a training set which features temperatures far outside the range of the experiment, but at a reduced occurrence, see figure~\ref{fig:training_data}. The training set consists of density profiles of condensates of various temperature, length, and atomnumber. 
More specifically, the temperature ranges from 10\,nK to 160\,nK in steps of 10\,nK, the number of atoms ranges from 4000 to 10000 in steps of 2000, and the length of the box-trap ranges from 50\,$\mu\mathrm{m}$ to 150\,$\mu\mathrm{m}$ in steps of 25\,$\mu\mathrm{m}$. For each combination of these three parameters we simulated 500 density profiles for temperatures close to the experimental range and fewer outside, see figure~\ref{fig:training_data_a}.

Having the extremum temperatures of the training data set as far away from the temperature region of interest as possible is crucial for reducing biases in the model. However, if too many hot realizations are present in the training set, their associated greater uncertainty will dominate the loss function, leading to less optimization of the performance in the relevant temperature range.
We found that employing the distribution of temperatures for the training set seen in figure~\ref{fig:training_data}, as opposed to a uniform distribution, greatly reduced the bias of our models.

\section{Benchmarking the neural network on simulated data} \label{app:benchmark}

\begin{figure*}
\center
\includegraphics[width =0.9\textwidth]{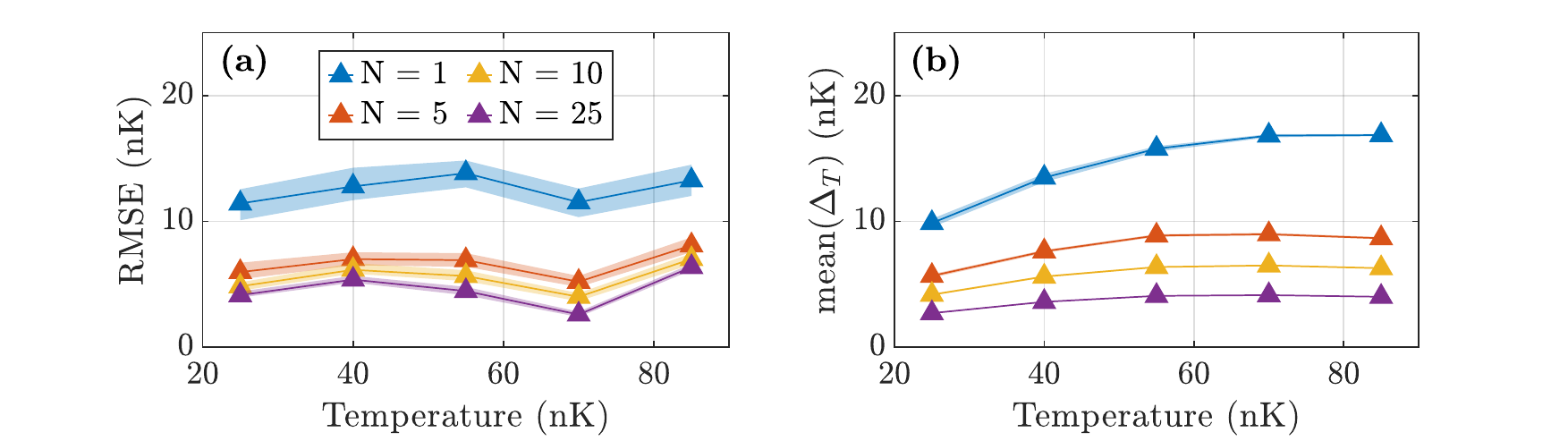}
\phantomsubfloat{\label{fig:benchmark_mean_a}}
\phantomsubfloat{\label{fig:benchmark_mean_b}}

\vspace{-1\baselineskip}

\caption{\label{fig:benchmark_mean} Benchmark of the neural network models predictions of the mean temperature $\bar{\mu}_T$ and the uncertainty thereof $\Delta_T$ for 11.2\,ms expansion time. The benchmark is performed on a simulated test data set with the same parameter distribution as the profiles used for figure~\ref{fig:training}.
(a) Root-mean-square error of the mean predictions. (b) Mean value of the uncertainties of the predictions. For each temperature, the results were generated by sampling $N$ density profiles, estimating their temperatures and computing the mean $\bar{\mu}_T$ and uncertainty $\Delta_T$, then repeating the process $M =250$ times and taking the average over the repetitions. The shaded areas show 95$\%$ confidence intervals obtained using bootstrapping (bias corrected accelerated method~\cite{10.1214/ss/1032280214}) with $M$ re-samplings.}
\end{figure*}

\subsection{Benchmark of predictions of the mean temperature}
In Section~\ref{sec:results}, we benchmarked the single realization predictions of the neural network models on a simulated test data set.
For a set of multiple density profiles, we can combine the temperature predictions of individual realizations to estimate the mean temperature of the sample and the uncertainty thereof, following eqs.~\eqref{eq:mean_estimate} and \eqref{eq:uncert_estimate}, respectively. In this section, we will repeat the benchmarking process described in Section~\ref{sec:results} but for combined predictions on sets $N = $ 1, 5, 10 and 25 density profiles at a time.

For the purpose of this benchmark, we will simulate a new test data set with the same parameter distribution as described in Section~\ref{sec:results}. Thus, for temperatures ranging from 25\,nK to 85\,nK in steps of 15\,nK, we simulate 200 realizations for each combination of atomnumbers 4000, 7000 and 10000 and box-trap lengths 50, 100 and 150\,$\mu\mathrm{m}$.
In total this yields 1800 different density profiles for each temperature. All profiles are simulated at  11.2\,ms time-of-flight.

Next, we sample subsets from the full test data set and employ the neural network model to predict the mean temperature of each subset $\bar{\mu}_T$ and the uncertainty thereof $\Delta_T$.
For each temperature and set size $N$, we sample $M=250$ times and averages will be computed over said samples. For practical reasons we sample the $N > 1$ subsets with replacement, meaning a single profile can feature multiple times in a single subset or in multiple subsets, due to limited size of the test data set. However, for the data set in question, we find that sampling with replacement has negligible effect on the benchmark quantities computed.

In figure~\ref{fig:benchmark_mean} we plot the results of the benchmark. The bias remains similar across all set sizes $N$ and is therefore not plotted. As $N$ increases, the estimate of the mean becomes increasingly better, as seen by the decreasing root-mean-square-error (RMSE) plotted in figure~\ref{fig:benchmark_mean_a}.  In particular, we observe a large increase in accuracy going from $N=1$ to $N=5$ profiles. Note, the RMSE is computed using the formula in main text, but with the single profile prediction $\mu_T$ replaced with the mean prediction $\bar{\mu}_T$.
Also the uncertainty decreases for all temperatures as the set size $N$ increases, which can be seen in figure~\ref{fig:benchmark_mean_b}.

\subsection{Benchmark with shot-to-shot variations}
In this section we benchmark the neural network model on simulated data emulating a realistic scenario, where both atom number and temperature of the condensate can vary between individual realizations. We generate a new set of test data: For a box of fixed length of 100\,$\mu\mathrm{m}$ we select a random number of atoms from a normal distribution with mean 6000 and a standard deviation of 300, and a random temperature from another normal distribution with mean 70\,nK and a standard deviation of 10\,nK. For each expansion time we generate a total of 1000 profiles.
Using this data set we will perform a statistical analysis using bootstrap techniques, similar to the analysis conducted for the experimental data set presented in Section~\ref{sec:results}.

\begin{figure*}
\center
\includegraphics[width =0.9\textwidth]{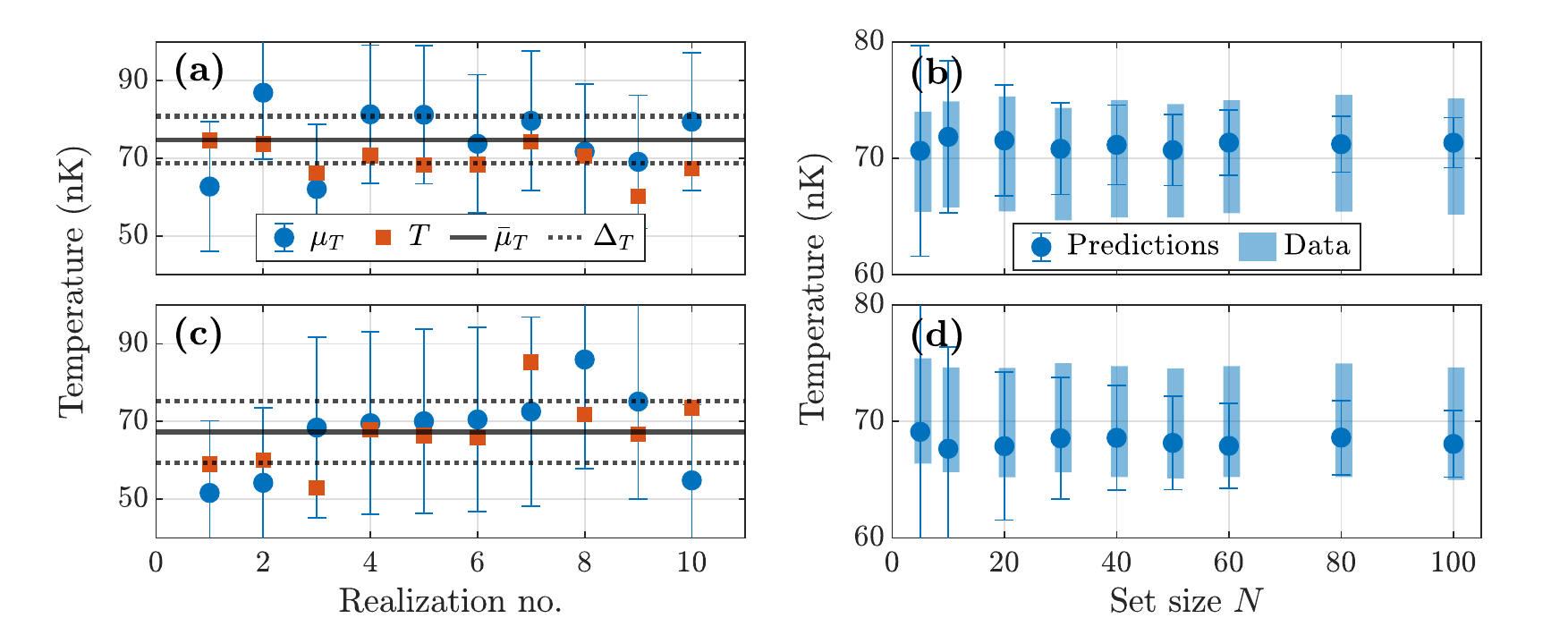}
\phantomsubfloat{\label{fig:benchmark_a}}
\phantomsubfloat{\label{fig:benchmark_b}}
\phantomsubfloat{\label{fig:benchmark_c}}
\phantomsubfloat{\label{fig:benchmark_d}}

\vspace{-1\baselineskip}

\caption{\label{fig:benchmark} Benchmark of both the individual and mean prediction of the neural network on simulated data.
Left column: Individual and mean prediction of the neural network on 10 profiles randomly drawn from the data set. The blue dots show the $\mu_T^{(i)}$ predictions of the model, while the error bars mark the uncertainties $\sigma_T^{(i)}$. For comparison, the actual temperatures of the analyzed density profiles are marked by orange squares.
Right column: Subsets of different sizes $N$ are sampled from the overall data set using bootstrapping techniques. For each $N$ the average of the mean temperature prediction $\bar{\mu}_T$ and uncertainty thereof $\Delta_T$ over the bootstrap samples is computed. The blue data points and error bars are said averages of $\bar{\mu}_T$ and $\Delta_T$, respectively. The shaded bars indicate the mean and standard deviation of the actual temperatures of the bootstrap sampled density profiles.
(a,b) TOF = 11.2\,ms. (c,d) TOF = 2\,ms.}
\end{figure*}

In figures \ref{fig:benchmark_a} and \ref{fig:benchmark_c} the individual predictions of the model on 10 randomly drawn profiles are shown, for 11.2\,ms and 2\,ms expansion time, respectively. We observe the true temperature of the condensate $T$ being within the uncertainty of the models prediction, thus demonstrating the models ability to accurately access their own uncertainty. Computing the mean temperature of the set $\bar{\mu}_T$ and the uncertainty thereof $\Delta_T$, we also find the true mean temperature of the sample within the corresponding uncertainty.

Next, subsets of various sizes $N$ are sampled using bootstrap techniques, and predictions of the subset mean temperatures are made. For each $N$ this is repeated 50 times, and the average of the predictions $\bar{\mu}_T$ and $\Delta_T$ is taken over bootstrap samples. The results are plotted in the figures \ref{fig:benchmark_b} and \ref{fig:benchmark_d}. For comparison, we also plot both the mean and standard deviation of the true temperatures of all the samples within a given set size. Note, the mean and standard variation of the samples vary slightly between each set size due to the finite number of samples. For small set sizes this is particularly obvious, whereas for large $N$ the number of samples becomes large enough that the mean and standard variation of the samples match the underlying distribution.

Here we find good agreement between the mean prediction of the models and the mean temperature of the samples. Even for relatively small set sizes the predictions of the neural networks are very consistent. Notably, the model trained for 2\,ms expansion time exhibits a slight negative bias, while the model trained for 11.2\,ms expansion time is slightly positively biased - both consistent with the benchmark of the models presented in figure~\ref{fig:training}.

\section{Generating artificial absorption images} \label{app:artificial}

The process of simulating images comprises of two main steps:
First, the state of an ultracold, trapped Bose gas is calculated. As discussed in the main text, the phase and density fluctuations of the condensate can be described by an Ornstein-Uhlenbeck process~\cite{PhysRevLett.105.015301}. Next, using the phase-density representation~\cite{PhysRevA.67.053615}, we obtain the wavefunction of a single realization. For each set of condensate variables (box length, atom number, temperature) we generate many individual realizations of the wavefunction, thereby sampling as many different fluctuations as possible. 
Finally, we simulate the experimental measurement process of the cold Bose gas, namely the free expansion of the gas followed by the absorption imaging. Note, the free expansion is only an approximation, as interactions between the atoms still take place in the early stages of the expansion, where the gas is still relatively dense. 
The simulation of the imaging process is described in great detail in Ref. \cite{schweigler2019correlations}. Hence, we will here simply summarize the various steps taken and the different effects accounted for. 

\subsubsection{Absorption of the imaging light.}
In the experiment we employ absorption imaging, meaning that for each measurement we take two images; one of the atomic cloud and a second image of only the imaging light. By comparing the measured light intensities of the two images, the light absorption of the atomic cloud, and thereby the atomic density integrated along the imaging direction, can be inferred using the Beer-Lambert law. If $x$ is the direction of imaging the Beer-Lambert law reads
\begin{equation}
\frac{\mathrm{d} I(x,y,z)}{\mathrm{~d} x}=-\sigma n(x, y, z) I(x,y,z) \; , \label{eq:BeerLambert}
\end{equation}
where $I$ is the intensity of the imaging light, $n$ is the atomic density, and $\sigma$ is the absorption cross section for the given transition. Upon switching on the imaging light, the populations of different $m_F$ states will rearrange themselves, eventually reaching a dynamic equilibrium. In our imaging system, a linear polarization along the quantization axis of the atoms is employed, causing the dynamic equilibrium hyperfine populations to form an effective two-level system. The transition in this effective two-level system has an effective absorption cross section, given by the on-resonance cross section $\sigma_0$ reduced by a factor of $\alpha = 0.54$~\cite{steck2001rubidium}. The lower absorption causes the saturation intensity to increase proportionally. Thus, the total absorption cross section reads
\begin{equation}
\sigma=\frac{\alpha \sigma_{0}}{1+\alpha \frac{I}{I_0^\mathrm{sat}}} \; . \label{eq:crosssec}
\end{equation}
For the transition used, the on-resonance saturation intensity is $I_0^\mathrm{sat}=16.6933 \mathrm{~W} / \mathrm{m}^{2}$, and the on-resonance cross section is $\sigma_0 = 2.905 \times 10^{-13}\, \mathrm{m}^2$.

By integrating the Beer-Lambert law of eq.~\eqref{eq:BeerLambert} along the imaging direction, one can solve for the integrated atomic density, yielding
\begin{equation}
\tilde{n}(y, z)=\frac{1}{\sigma_{0}}\left[-\frac{1}{\alpha} \ln \left(\frac{I_{a} (y,z)}{I_{0}(y,z)}\right)+\frac{I_{0}(y,z)-I_{a}(y,z)}{I_0^\mathrm{sat}}\right] \; .
\end{equation}
Here, $I_0$ and $I_a$ are the intensity profiles of the imaging light before and after having passed through the atomic cloud, respectively. In practice, $I_a$ is obtained from the picture of the cloud, while $I_0$ is the picture without the cloud present.
For the purpose of simulating images, the Beer-Lambert law can also be used to compute the intensity profile of the imaging light after having passed through the atomic cloud
\begin{equation}
I_{a}( y, z)=\frac{I_0^\mathrm{sat}}{\alpha} W\left[\frac{\alpha I_{0}( y,z)}{I_0^\mathrm{sat}} \; e^{\frac{\alpha I_{0}( y, z)}{I_0^\mathrm{sat}}-\alpha \tilde{n}(y, z)  \sigma_{0}} \right] \; ,
\end{equation}
where $W$ denotes the Lambert-W function on its principal branch.

\subsubsection{Recoil blurring.}
During the imaging process, the atoms absorb photons from the imaging light and have a probability of spontaneously re-emitting them. Since the direction of these re-emitted photons is random, their recoil causes the atom to move following a random walk leading to blurring of the image. The process above causes a continuous deformation of the cloud throughout the exposure time of the imaging. We estimate the blurring by discretizing the deformation and thus subdividing the exposure time into small time slices $\Delta t$. If we let $N_{abs}$ be the number of photons absorbed by a single atom during the time $\Delta t$, the expectation value of the square distances travelled in said duration is~\cite{ketterle1999making} 
\begin{equation}
\left\langle|\vec{r}|^{2}\right\rangle_{\mathrm{abs}} \approx v_{\mathrm{rec}}^{2} \Delta t^{2} \frac{N_{abs}}{3} \; ,
\end{equation}
where $v_{\mathrm{rec}}$ is the recoil velocity. The blurring of the cloud can then be estimated by convolving the 2d atomic density with a Gaussian of width 
\begin{equation}
\sigma_{\mathrm{rec}}^{2} = \frac{1}{3} \left\langle|\vec{r}|^{2}\right\rangle_{\mathrm{abs}}  \; .
\end{equation}

\subsubsection{Imaging resolution of an expanded cloud.}
In order to simulate the image formation using a coherent light source we need to compute the coherent transfer function $c\left(k_{x}, k_{y}\right)$, which relates the image in the object plane to the light field amplitude in the object plane, i.e.~the light field immediately after having passed through the atomic cloud.
The coherent transfer function also captures the resolution of the imaging system, which, in part, is determined by the numerical aperture (NA). A higher NA means better in-focus resolution but worse depth of focus. Thus, when imaging an extended object, such as the condensate after free expansion, the optical resolution can be substantially reduced. 
Let us assume the density of the atomic cloud after free expansion and along the imaging direction to be a Gaussian of width $w$. Assuming a semi-transparent cloud and a circular aperture, the effective coherent transfer function reads~\cite{B_cker_2009}
\begin{equation}
c_{e f f}\left(k_{y}, k_{z}\right) \propto \Theta\left(\frac{\mathrm{NA}}{\lambda}-k_{t}\right) \mathrm{e}^{-\left(\pi \frac{\lambda}{\sqrt{2}} k_{t}^{2} w\right)^{2}} \mathrm{e}^{-i \pi \lambda k_{t}^{2} x_{0}} \; , \label{eq:CTFeff}
\end{equation}
where $k_{t}^{2}=k_{y}^{2}+k_{z}^{2}$, $x_{0}$ is the distance from the center of the cloud to the imaging focus plane, and $\lambda = 780\, \mathrm{nm}$ is the wavelength of the imaging light. $\Theta$ denotes the Heaviside function. The first term in eq.~\eqref{eq:CTFeff} represents the fundamental diffraction limit of the optical system, while the second term is the decrease in resolution due to the finite extend of the Gaussian cloud.

\subsubsection{Properties of the camera.}
Lastly, we account for the properties of our camera. First, the intensity profiles of the imaging light are coarse grained by binning them according to the camera pixels, which have a length of $1.05 \, \mu\mathrm{m}$ in the object plane. Next, the quantum efficiency of the camera is accounted for. Finally, we account for the shot noise on images by assuming the arrival of photons at every pixel to follow a Poisson distribution.

\bibliography{references}

\end{document}